\documentclass[sigplan,screen,authorversion]{acmart}
\AtBeginDocument{%
  \providecommand\BibTeX{{%
    \normalfont B\kern-0.5em{\scshape i\kern-0.25em b}\kern-0.8em\TeX}}}

\usepackage{prelude}

\setcopyright{acmlicensed}
\acmPrice{15.00}
\acmDOI{10.1145/3609025.3609477}
\acmYear{2023}
\copyrightyear{2023}
\acmSubmissionID{icfpws23funarchmain-id3-p}
\acmISBN{979-8-4007-0297-6/23/09}
\acmConference[FUNARCH '23]{Proceedings of the 1st ACM SIGPLAN International Workshop on Functional Software Architecture}{September 8, 2023}{Seattle, WA, USA}
\acmBooktitle{Proceedings of the 1st ACM SIGPLAN International Workshop on Functional Software Architecture (FUNARCH '23), September 8, 2023, Seattle, WA, USA}
\received{2023-06-01}
\received[accepted]{2023-06-28}

\begin{document}

\title{Typed Design Patterns for the Functional Era}

\author{Will Crichton}
\email{wcrichto@brown.edu}
\orcid{0000-0001-8639-6541}
\affiliation{
  \institution{Brown University}           
  \city{Providence}
  \state{Rhode Island}
  \postcode{02912}
  \country{USA}                    
}


\begin{abstract}
    This paper explores how design patterns could be revisited in the era of mainstream functional programming languages. I discuss the kinds of knowledge that ought to be represented as functional design patterns: architectural concepts that are relatively self-contained, but whose entirety cannot be represented as a language-level abstraction. I present four concrete examples embodying this idea: the Witness, the State Machine, the Parallel Lists, and the Registry. Each pattern is implemented in Rust to demonstrate how careful use of a sophisticated type system can better model each domain construct and thereby catch user mistakes at compile-time.
\end{abstract}

\ccsdesc[500]{Software and its engineering~Design patterns}
\ccsdesc[500]{Software and its engineering~Functional languages}

\keywords{design patterns, domain-driven design, rust}



\maketitle

\section{Introduction}

\emph{Where are all the functional programming design patterns?} People have been clamoring for answers to this question for over a decade\,\citeyearpar{where-are-the-patterns}. As functional programming concepts have crept into the mainstream of software engineering, that clamor has only grown. Yet, no catalog has emerged as the clear functional successor to the venerable \textit{Design Patterns}\,\cite{gof}. This paper explores the question: what might a catalog of functional design patterns look like?

First, it is worth discussing the definition and purpose of a software design pattern. \citet{norvig1998dyn} argued that a design pattern should provide ``descriptions of what experienced designers know
(that isn't written down in the Language Manual)''. The Gang of Four (GoF) further explain:
\begin{quote}
    ``Each design pattern systematically names, explains, and evaluates an
important and recurring design in object-oriented systems. Our goal is to capture design experience in a form that people can use effectively. To this end we have documented some of the most important design patterns and present them as a catalog.''\,\cite[p.~4]{gof}
\end{quote}

\noindent Put another way, a design pattern satisfies two key criteria:
\begin{enumerate}
    \item It captures a recurring phenomenon (a pattern) in real-world system design.
    \item It provides a software design strategy for managing this phenomenon that \textit{is difficult to describe mechanistically.}
\end{enumerate}

The first criterion points to why functional design patterns are still uncommon. There are not nearly as many large-scale systems written with functional languages compared to OOP languages. But this is slowly changing --- the last decade has seen an explosion of systems implemented in functional languages\footnotemark{} like Rust, Scala, Swift, and Clojure. I personally have close to a decade of experience with Rust, which underlies the perspective given in this paper. After working with Rust systems for web development, 3D graphics, data analytics, etc., I have seen many patterns of system design with various encodings into Rust. But I have never seen these patterns articulated concisely and collated into a single catalog.

\footnotetext{This list may raise some eyebrows as to my definition of ``functional''. I mean languages with features associated with the intellectual lineage of the lambda calculus.}

The second criterion points to why the idea of design patterns has been tricky to translate into a functional setting. A common critique of the GoF patterns is that their complexity stems from an impoverished implementation language like C++.  \citet{norvig1998dyn} showed that use of a dynamically-typed language like Lisp or Dylan can dramatically simplify some patterns to the point of triviality. \citet{gibbons2006dp} showed that the use of a Haskell-esque type system enables patterns to be directly represented as well-typed library-level abstractions. From this perspective, one could argue that the Haskell or Rust standard libraries are design pattern catalogues \textit{per se}, just communicated via code rather than prose.

However, it is undoubtedly true that an FP novice cannot simply read the Haskell or Rust language manuals and proceed to effectively design a complex system. A key skill that bridges the gap is understanding how to \emph{systematically map domain concepts to language features}. With object-oriented programming, developers have the problem of modeling a diverse range of domain concepts with a small set of load-bearing features.
By contrast, functional languages offer a dizzying array of modeling tools: products, sums, newtypes, generics, modules, traits, type families, substructural types, lifetimes, and so on. FP novices therefore have a different problem: deciding which combination of many features is most appropriate for modeling a domain concept.

This paper articulates a vision for (typed) functional design patterns to match this new reality. I propose that functional design patterns should articulate concrete mappings from common domain concepts to particular configurations of functional language features. That vision is presented through four concrete patterns: the Witness, the State Machine, the Parallel Lists, and the Registry. The selection, structure, and substance of each pattern embodies my argument for what constitutes a useful software design pattern.

\section{A Sample of Functional Design Patterns}
\label{sec:patterns}
\newenvironment{patpart}[1]{\noindent\textbf{#1:}}{\vspace{0.3em}}

Drawing inspiration from domain-driven design\,\cite{evans2003ddd,wlaschin2017ddd}, each design pattern is centered around a ``domain pattern'', i.e. a language-independent concept that appears in many software application domains. A domain pattern is described in terms of a ``schema'', in the sense used in psychological schema theory\,\cite{rumelhart1980schemata}.  The knowledge captured in each design pattern is about how to model the domain pattern at both the type-level and expression-level of a particular functional programming language. The quality of a model is evidenced by the fact that incorrect uses of the model cause type errors rather than runtime errors (or worse, undefined behavior). Each pattern is described using the following structure:

\begin{enumerate}
\item \textbf{Schema:} the abstract elements (written in \pat{Small Caps}) and relations that characterize the domain pattern.
\item \textbf{Examples:} a short list of instances of the pattern in real-world applications, briefly fit to the schema.
\item \textbf{Case study:} an extended case study of how to implement one instance of the pattern.
\item \textbf{Commentary:} some additional notes about the pattern.
\end{enumerate}

The case studies are all implemented in Rust. That decision is in part due to my familiarity with the language. But it bears emphasizing that Rust is an excellent medium for articulating functional design patterns. Rust is an exceptionally practical language, designed for implementing production-grade systems rather than as a vehicle for research. Functional design patterns should focus on mappings that are \emph{practical} as much as \emph{feasible}, and the best way to determine practicality is under heavy stress from real-world use.

\newpage

\subsection{Witness}

\begin{patpart}{Schema} 
An \pat{Action} that can only execute once a particular \pat{Condition} is satisfied.
\end{patpart}

\begin{patpart}{Examples}
Access control (e.g., a user must login before seeing their profile), resource management (e.g., a computation can only occur if the machine has enough resources).
\end{patpart}

\begin{patpart}{Case study}
Consider a website that has normal and admin users, and an admin panel that should only be accessed (the \pat{Action}) to users logged in as an admin (the \pat{Condition}). The base API contains the following methods:

\begin{itemize}
    \item \mintinline{Rust}|render_admin_panel() -> Html|: returns the HTML for the admin panel.
    \item \mintinline{Rust}|render_404() -> Html|: returns the HTML for a 404 page.
    \item \mintinline{Rust}|current_user() -> User|: returns the currently logged-in user.    
    \item \mintinline{Rust}|is_admin(&User) -> bool|: returns true if the user is an admin.
\end{itemize}

\Cref{lst:admin-flag} provides both a correct and incorrect (but well-typed) implementation of a route for the admin panel. The correct admin panel route verifies that the current user is an admin, and returns a 404 otherwise. The issue with this implementation is that if we forget to check that the current user is an admin, or e.g. mess up the check, then the program could permit non-admins to see the admin panel.

\begin{listing}[h]
\begin{minted}{rust}
#[route("/admin")]
fn admin_panel() -> Html {
  if is_admin(&current_user()) {
    render_admin_panel();
  } else {
    render_404();
  }
}

#[route("/admin")]
fn admin_panel_bad() -> Html {
  // Allowed to compile, leaks information
  render_admin_panel()
}
\end{minted}
\vspace{-1em}
\caption{Admin panel with a boolean encoding of the admin flag on users.}
\label{lst:admin-flag}
\end{listing}

An alternative encoding is shown in \Cref{lst:admin-witness}. The key idea is to create a new data type that ``witnesses'' the proof of a capability, i.e. that a user is an admin. The \mintinline{rust}|Admin| data type is wrapped in a module so it can only be constructed via the \mintinline{rust}|try_admin| method. The \mintinline{rust}|render_admin_panel| function is modified such that it takes an \mintinline{rust}|Admin| as input. Therefore the only well-typed way to call the function is to generate a proof that the user (or at least \textit{a} user) is an admin, and pass that proof to the renderer.

\begin{listing}[h]
\begin{minted}{rust}
mod admin {
  pub struct Admin {}
  impl User {
    pub fn try_admin(&self) -> Option<Admin> {
      if is_admin(self) { Some(Admin {}) }
      else { None }
    }
  }
}    

fn render_admin_panel(_admin: Admin) -> Html;

#[route("/admin")]
fn admin_panel() -> Html {
  if let Some(admin) = current_user().try_admin() {
    render_admin_panel(admin);
  } else {
    render_404();
  }
}

#[route("/admin"])
fn admin_panel_bad() -> Html {
   // Doesn't compile, missing witness
   render_admin_panel()
   // Doesn't compile, can't manually build witness
   render_admin_panel(Admin {})
}
\end{minted}
\vspace{-1em}
\caption{Admin panel with a witness encoding of the admin permission.}
\label{lst:admin-witness}
\end{listing}

\end{patpart}

\begin{patpart}{Commentary}
While many type-driven design techniques are inspired by theorem proving, the witness technique is most directly drawn from that intellectual lineage. As applications become more security- and privacy-sensitive, design patterns can help developers write correct-by-construction code with respect to issues like access control.
This pattern has been adopted in Rust web applications, most notably Rocket\,\citeurl{https://rocket.rs/}, as well as some cryptography libraries.
\end{patpart}

\subsection{State Machine}

\begin{patpart}{Schema} 
An object that exists in one of multiple \pat{States}, which can \pat{Transition} to other \pat{States} in response to \pat{Events}.\end{patpart}

\begin{patpart}{Examples} 
Files (open and closed), mutexes (unlocked and locked), shopping carts (empty, filled, purchased).
\end{patpart}

\begin{patpart}{Case study}
Consider a file model with three \pat{States}: reading, end-of-file, and closed. \Cref{fig:file-sm} shows a diagram of the model.
\begin{figure}[h!]
\begin{center}
\begin{tikzpicture}[scale=0.17]
\tikzstyle{every node}+=[inner sep=0pt]
\draw [black] (19.2,-31) circle (3);
\draw (19.2,-31) node {\small Read};
\draw [black] (35.3,-31) circle (3);
\draw (35.3,-31) node {\small Eof};
\draw [black] (26.9,-42) circle (3);
\draw (26.9,-42) node {\small Close};
\draw [black] (26.9,-42) circle (2.4);
\draw [black] (20.92,-33.46) -- (25.18,-39.54);
\fill [black] (25.18,-39.54) -- (25.13,-38.6) -- (24.31,-39.17);
\draw (22.45,-37.86) node [left] {\mintinline{rust}|sys_close|};
\draw [black] (33.48,-33.38) -- (28.72,-39.62);
\fill [black] (28.72,-39.62) -- (29.6,-39.28) -- (28.81,-38.68);
\draw (31.67,-37.9) node [right] {\mintinline{rust}|sys_close|};
\draw [black] (22.2,-31) -- (32.3,-31);
\fill [black] (32.3,-31) -- (31.5,-30.5) -- (31.5,-31.5);
\draw (27.25,-30.5) node [above] {\mintinline{rust}|sys_read|};
\draw [black] (17.877,-28.32) arc (234:-54:2.25);
\draw (19.2,-23.75) node [above] {\mintinline{rust}|sys_read|};
\fill [black] (20.52,-28.32) -- (21.4,-27.97) -- (20.59,-27.38);
\draw [black] (14,-31) -- (16.2,-31);
\draw (13.5,-31) node [left] {\mintinline{rust}|sys_open|};
\fill [black] (16.2,-31) -- (15.4,-30.5) -- (15.4,-31.5);
\end{tikzpicture}
\end{center}
\vspace{-1em}
\caption{Model of the file state machine.}
\label{fig:file-sm}
\end{figure}
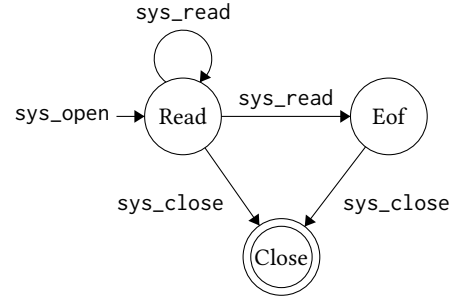
An operating system provides a low-level API (i.e. a set of \pat{Transitions}) that implements the model:
\begin{itemize}
    \item \mintinline{Rust}|sys_open(&str) -> Descriptor|: creates a file descriptor. 
    \item \mintinline{Rust}|sys_eof(&Descriptor) -> bool|: checks if the file has reached EOF, or panics if the file is closed.
    \item \mintinline{Rust}|sys_read(&mut Descriptor) -> Vec<u8>|: returns a data buffer from the file, or panics if the file has reached EOF or closed. 
    \item \mintinline{Rust}|sys_close(&mut Descriptor)|: closes the file, or panics if the file is already closed.
\end{itemize}

Consider the task of designing a \mintinline{rust}|File| type that encapsulates the low-level API. \Cref{lst:fs-wrap} shows a trivial wrapper. This wrapper provides no protection against incorrect usage. For example, one could attempt to read a file after reaching EOF which would panic, such as in \Cref{lst:fs-wrap-ex}.

\begin{listing}[h]
\begin{minted}{rust}
struct File { fd: Descriptor }
impl File {
  fn open(path: &str) -> File {
    File { fd: sys_open(path) }
  }
  fn eof(&self) -> bool {
    sys_eof(&self.fd)
  }
  fn read(&mut self) -> Vec<u8> {
    sys_read(&mut self.fd)
  }
  fn close(&mut self) {
    sys_close(&mut self.fd);
  }
}
\end{minted}
\vspace{-1em}
\caption{File state machine that directly wraps the system API.}
\label{lst:fs-wrap}
\end{listing}

\vspace{-1em}

\begin{listing}[h]
\begin{minted}{rust}
fn main() {
  let mut f = File::open("f.txt");
  // permits reading past EOF
  f.read(); f.read(); f.read();
  // permits closing multiple times
  f.close(); f.close();
  // permits reading after close
  f.read();
}
\end{minted}
\vspace{-1em}
\caption{Example usage of the file state machine wrapper API.}
\label{lst:fs-wrap-ex}
\end{listing}
\vspace{-0.5em}

One way to make the API less error-prone is to represent all states within one enumerated type, and then change the methods to return option types that indicate incorrect state transitions. \Cref{lst:fs-enum} shows an implementation of this idea. The enum representation ensures that panics cannot happen within the \mintinline{rust}|File| implementation. However, a user may still encounter unexpected \mintinline{rust}|None| values at runtime from an incorrect usage of the API as shown in \Cref{lst:fs-enum-ex}.

\begin{listing}[h]
\begin{minted}{rust}
enum State { Read, Eof, Close }
struct File { fd: Descriptor, state: State }

impl File {
  fn open(path: &str) -> File {
    File { fd: sys_open(path), state: State::Read }
  }

  fn read(&mut self) -> Option<Vec<u8>> {
    match self.state {
      State::Read => {
        let buf = sys_read(&mut self.fd);
        if sys_eof(&self.fd) { 
          self.state = State::Eof; 
        }
        Some(buf)
      }
      State::Eof | State::Close => None
    }
  }

  fn close(&mut self) -> Option<()> {
    match self.state {
      State::Read | State::Eof => {      
        sys_close(&mut self.fd);
        self.state = State::Close;
        Some(())
      }
      State::Close => None
    }
  }
}
\end{minted}
\vspace{-1em}
\caption{File state machine with enum encoding.}
\label{lst:fs-enum}
\end{listing}
\vspace{-1em}
\begin{listing}[h]
\begin{minted}{rust}
fn main() {
  let mut f = File::open("hello.txt");
  while let Some(data) = f.read() {
    // process the data
  }
  f.close().unwrap();
  // can still get a runtime error,
  // but represented as a None
  f.read().unwrap();
  f.close().unwrap();   
}
\end{minted}
\vspace{-1em}
\caption{Example of the enum-based file state machine.}
\label{lst:fs-enum-ex}
\end{listing}

These runtime panics can be turned into compile-time errors by encoding each state as a distinct type, and then only attaching the appropriate methods to each type. \Cref{lst:fs-typestate} shows one such implementation with \mintinline{rust}|FileRead| and \mintinline{rust}|FileEof| states. The other key idea is that each state transition method invalidates the input object and returns the new state as a part of the output. In Rust, this concept is represented by taking ownership of the input state via  \mintinline{rust}|self| as opposed to a reference via \mintinline{rust}|&self| or \mintinline{rust}|&mut self|. 

For example, the signature of \mintinline{rust}|read| consumes the input file, and then uses an \mintinline{rust}|Either| type to represent that one of two state transitions is possible. Either the file self-transitions to the Read state, or the file transitions to the Eof state. For the \mintinline{rust}|close| method, no state is returned because because no operations exist on the Close state. 

\begin{listing}[h]
\begin{minted}{rust}
struct FileRead { fd: Descriptor }
struct FileEof { fd: Descriptor }
enum Either<L, R> { Left(L), Right(R) }

impl FileRead {
  fn open(path: &str) -> FileRead {
    FileRead { fd: sys_open(path) }
  }

  fn read(mut self) -> 
    Either<(FileRead, Vec<u8>), FileEof> 
  {
    if sys_eof(&self.fd) {
      Either::Right(FileEof { fd: self.fd })
    } else {
      let buf = sys_read(&mut self.fd);
      Either::Left((self, buf))
    }
  }

  fn close(mut self) {
    sys_close(&mut self.fd);
  }
}

impl FileEof {
  fn close(mut self) {
    sys_close(&mut self.fd);
  }
}
\end{minted}
\vspace{-1em}
\caption{File state machine with typestate encoding.}
\label{lst:fs-typestate}
\end{listing}

\vspace{-1em}

\begin{listing}[h]
\begin{minted}{rust}
fn main() {
  let mut f: FileRead = FileRead::open("f.txt");
  let f: FileEof = loop {
    f = match f.read() {
      Either::Left((f, data)) => {
        // ... process the data ...
        f
      }
      Either::Right(f) => break f
    }
  };
  // cannot call f.read() on FileEof
  f.close();
  // cannot call f.close() again since f is moved
}
\end{minted}
\vspace{-1em}
\caption{Example of the typestate-based file state machine.}
\label{lst:fs-typestate-ex}
\end{listing}

A drawback of the approach in \Cref{lst:fs-typestate} is that both struct fields like \mintinline{rust}|fd| and methods like \mintinline{rust}|close| are duplicated between the two state structs.  \Cref{lst:fs-generic} shows an alternative approach. We return to having a single \mintinline{rust}|File| type, but now it is parameterized\footnotemark{} by a state type \mintinline{rust}|S|. The implementation is similar to \Cref{lst:fs-typestate}, except every instance of \mintinline{rust}|FileRead| is replaced by \mintinline{rust}|File<Read>|. Additionally, the \mintinline{rust}|close| method is now implemented once using a generic \mintinline{rust}|impl| block over all states.

\footnotetext{
  The \mintinline[fontsize=\footnotesize]{rust}|File<S>| definition needs a field \mintinline[fontsize=\footnotesize]{rust}|PhantomData<S>| to prevent the compiler from complaining about an unused type, but we omit that detail here.
}

\begin{listing}[h]
\begin{minted}{rust}
struct Read;
struct Eof;
struct File<S> { fd: Descriptor }

impl File<Read> {
  fn open(path: &str) -> File<Read> {
    /* Essentailly same implementation */
  }

  fn read(mut self) -> 
    Either<(File<Read>, Vec<u8>), File<Eof>> 
  {
    /* Essentially same implementation */
  }
}

impl<S> File<S> {
  fn close(mut self) {
    sys_close(&mut self.fd);
  }
}
\end{minted}
\vspace{-1em}
\caption{File state machine with generic typestate encoding.}
\label{lst:fs-generic}
\end{listing}
\end{patpart}

\begin{patpart}{Commentary}
State machines are an extremely common idiom in system design, so it is a good instance of a kind of domain pattern that deserves treatment as a design pattern. The typestate technique has a long history\,\cite{strom1986typestate,aldrich2009typestate}, and recent work has shown how typestate can be embedded in the type systems of existing languages\,\cite{bernady2017linearhaskell,duarte2021rustts}. Libraries for session types use similar mechanisms\,\cite{pucella2008session,jespersen2015sessiontypes}.

In the Rust ecosystem, typestate can be found in hardware programming, e.g. to represent the state of GPIO pins on a board as in the rppal library\,\citeurlwith{https://github.com/golemparts/rppal}{golemparts/rppal} for Raspberry Pi. Typestate is also used by the Rocket web server framework\,\citeurl{https://rocket.rs/} to represent the different states of a server while it is starting up.
\end{patpart}

\subsection{Parallel Lists}

\begin{patpart}{Schema}
Two \pat{Lists} of heterogeneous elements with a \pat{Parallel Relation} between the elements.
\end{patpart}

\begin{patpart}{Examples}
\mintinline{rust}|printf| (string format holes should match the arguments to fill the holes), web routes (a URL with multiple parameters should match the server callback receiving those parameters).
\end{patpart}

\begin{patpart}{Case study}
Consider a \mintinline{rust}|printf|-style string formatter which takes a template \pat{List} (string literals and holes) and an argument \pat{List} (values to stringify into the holes). \Cref{lst:printf-vec} shows a simple implementation of printf (ignoring template parsing).

\begin{listing}[h]
\begin{minted}{rust}
enum FElem {
  Str(String),
  Arg
}

fn format(
  tpl: Vec<FElem>, mut args: Vec<&dyn ToString>
) -> String {
  tpl.into_iter()
    .map(|elem| match elem {
      FElem::Str(s) => s,
      FElem::Arg => args.remove(0).to_string(),
    })
    .collect()
}
\end{minted}
\vspace{-1em}
\caption{Formatting with enum/vector encoding.}
\label{lst:printf-vec}
\end{listing}
\vspace{-1em}
\begin{listing}[h]
\begin{minted}{rust}
fn main(){
  // Correct usage
  let tpl = vec![
    FElem::Str("Hello ".into()), FElem::Arg];
  let args = vec![&"World"];
  assert_eq!("Hello World", format(tpl, args));

  // Panics with too few arguments
  format(
    vec![FElem::Str("Hello ".into()), FElem::Arg],
    vec![]
  );

  // Silently ignores too many arguments
  format(
    vec![FElem::Str("Hello ".into()), FElem::Arg],
    vec![&"World", &"Again"]
  );
}
\end{minted}
\vspace{-1em}
\caption{Example of enum/vector-encoded formatter.}
\label{lst:printf-vec-ex}
\end{listing}

An API client can use this interface incorrectly which results in runtime errors. \Cref{lst:printf-vec-ex} shows how too few arguments will cause a runtime panic (because \mintinline{rust}|Vec::remove| will panic if there are no elements in the vector). More insidiously, extra arguments are simply ignored.

The key observation is that printf has a \pat{Parallel Relation} between the template and arguments. The number of \mintinline{rust}|Var| holes should match the size of the \mintinline{rust}|args| vector. (A more sophisticated model might also specify that type-specific formatting modifiers like \mintinline{rust}|".1f"| should only be used for floats.) 

To catch user mistakes at compile-time, the type system must be aware of the length and type of elements in each list. One technique for this is the \textit{heterogeneous list}, or h-list. An h-list is a list of elements with mixed types whose size is known at compile-time. \Cref{lst:printf-hlist} shows how to implement a type-safe formatter using h-lists.

\begin{listing}[h]
\begin{minted}{rust}
struct FStr(String);
struct FArg;

struct HNil;
struct HCons<H, T> { head: H, tail: T }

trait Format<Args> {
  fn format(&self, args: Args) -> String;
}

impl Format<HNil> for HNil {
  fn format(&self, _args: HNil) -> String {
    "".to_string()
  }
}

impl<ArgList, FmtList> Format<ArgList>
for HCons<FStr, FmtList>
where FmtList: Format<ArgList>
{
  fn format(&self, args: ArgList) -> String {
    let HCons { head: FStr(head), tail } = self;
    head.to_string() + &tail.format(args)
  }
}

impl<T, ArgList, FmtList> Format<HCons<T, ArgList>>
for HCons<FArg, FmtList>
where FmtList: Format<ArgList>, T: ToString,
{
  fn format(&self, args: HCons<T, ArgList>) 
    -> String 
  {
    args.head.to_string() + 
      &self.tail.format(args.tail)
  }
}
\end{minted}
\vspace{-1em}
\caption{Printf with h-list encoding.}
\label{lst:printf-hlist}
\end{listing}

\begin{listing}[t]
\begin{minted}{rust}
fn main() {
  let tpl = hlist![FStr("Hello ".into()), FArg];
  let args = hlist!["World"];
  assert_eq!("Hello World", tpl.format(args));

  // Type error with too few args
  tpl.format(hlist![])

  // Type error with too many args
  hlist![].format(hlist!["extra"])
}
\end{minted}
\vspace{-1em}
\caption{Example of h-list-encoded printf.}
\label{lst:printf-hlist-ex}
\end{listing}

The core h-list uses an inductive cons-list representation of \mintinline{rust}|HNil| and \mintinline{rust}|HCons<H, T>| types.
For example, lines 2-3 of \Cref{lst:printf-hlist-ex} show the creation of h-lists through an \mintinline{rust}|hlist!| macro. The type of \mintinline{rust}|tpl| is \mintinline{rust}|HCons<FStr, HCons<FArg, HNil>>|.
The formatting function is then expressed as an inductive computation over the template and argument h-lists. Unfortunately, the implementation is far less straightforward than the enum/vector encoding in \Cref{lst:printf-vec}. In short, the computation is defined through the \mintinline{rust}|format| method wrapped in the \mintinline{rust}|Format| trait. This trait is implemented for template h-lists, i.e. h-lists whose head types are either \mintinline{rust}|FArg| or \mintinline{rust}|FStr|. The base case of \mintinline{rust}|HNil| is the empty string (lines 11-15). The inductive case of \mintinline{rust}|HCons<FStr, ...>| (lines 17-25) extracts the inner string from \mintinline{rust}|FStr| and recurses. 

The key is the inductive case of \mintinline{rust}|HCons<FArg, ...>| (lines 27-37) which captures the parallelism between the template and argument lists. The \mintinline{rust}|Format| trait is only implemented when the argument list is an \mintinline{rust}|HCons<T, ...>| where \mintinline{rust}|T: ToString|. The implementation converts \mintinline{rust}|T| to a string, and then recurses on \textit{both} \mintinline{rust}|self.tail| (the format list) and \mintinline{rust}|args.tail| (the argument list). 
The resulting design catches the issues in \Cref{lst:printf-vec-ex} at compile-time, as shown in \Cref{lst:printf-hlist-ex}. Passing both too many and too few arguments results in a type error.
\end{patpart}

\begin{patpart}{Commentary}
The Parallel List pattern is a good example of the trade-off between complexity and safety. The enum/vector encoding is simpler to read for the implementor, but it can lead to more bugs at runtime. However, although the h-list approach catches bugs, the type errors can potentially be incomprehensible, especially to novices.

\citet{kiselyov2004hlist} pioneered the h-list approach for Haskell, which has since been adopted by libraries such as Servant\,\cite{mestanogullari2015servant}. H-lists are used in Rust by the web server library Warp \citegithub{https://github.com/seanmonstar/warp} and a popular MySQL client\,\citegithub{https://github.com/blackbeam/rust-mysql-simple}.
Many Rust libraries like Diesel\,\citeurl{https://diesel.rs/} and Bevy\,\citeurl{https://bevyengine.org/} emulate h-lists by implementing a trait for all tuples up to some size $N$, usually $N = 16$ or $32$. However, this approach seriously reduces the comprehensibility of type errors\,\citeurlwith{https://github.com/rust-lang/rfcs/pull/2397}{rust-lang/rfcs\#2397}. Yet another approach would be language-level variadic generics, but that feature is unlikely to be implemented in Rust soon\,\citeurlwith{https://github.com/rust-lang/rfcs/issues/376}{rust-lang/rfcs\#376}.
\end{patpart}

\subsection{Registry}
\begin{patpart}{Schema} 
Objects that map \pat{Keys} to heterogeneous \pat{Values}, and users can register \pat{Keyed Requests} for \pat{Values}.
\end{patpart}

\begin{patpart}{Examples} 
Event systems (callbacks associated with event names), dependency injection (named component injected into functions requesting components by name). 
\end{patpart}

\begin{patpart}{Case study}
Consider an event system that contains events (the \pat{Value}) with names (the \pat{Key}), such as \mintinline{rust}|OnClick| to represent a mouse click. Clients can register callbacks for an event (the \pat{Keyed Request}) which can be triggered with an event of the same name. The system should be open-world---the set of events can be extended by the API client.

A straightforward encoding of this model (e.g., as you can find in Javascript's DOM API) would represent event names as strings and event listeners as functions consuming a dynamically-typed event payload. \Cref{lst:event-string} shows the Rust implementation of this model.

\begin{listing}[h!]
\begin{minted}{rust}
type Listener = Box<dyn Fn(&dyn Any)>;

#[derive(Default)]
struct Events {
  listeners: HashMap<String, Vec<Listener>>
}

impl Events {
  fn register(&mut self, ev: &str, f: Listener) {
    self.listeners.entry(ev.into())
      .or_default().push(f);
  }

  fn trigger(&self, ev: &str, data: &dyn Any) {
    if let Some(fs) = self.listeners.get(ev) {
      for f in fs { f(data); }
    }
  }
}
\end{minted}
\vspace{-1em}
\caption{Event registry with string-key encoding.}
\label{lst:event-string}
\end{listing}
\vspace{-1em}
\begin{listing}[h!]
\begin{minted}{rust}
struct OnClick { mouse_x: usize, mouse_y: usize }
let mut events = Events::default();

// Correct example of register and trigger usage
events.register("click", Box::new(|ev| {
  let ev = ev.downcast_ref::<OnClick>().unwrap();
  assert_eq!(ev.mouse_x, 1);
}));
let ev = OnClick { mouse_x: 1, mouse_y: 3 };
events.trigger("click", &ev);

// Error #1: Wrong event name at register time
events.register("clack", ...);

// Error #2: Wrong event type in callback
events.register("click", Box::new(|ev| {
  let ev = ev.downcast_ref::<OnKeyPress>().unwrap();
}));

// Error #3: Wrong event name at trigger time
events.trigger("clack", ...);

// Error #4: Wrong event type passed to trigger
events.trigger("click", &OnKeyPress { /* ... */ });
\end{minted}
\vspace{-1em}
\caption{Example usage of the string-key event encoding.}
\label{lst:event-string-ex}
\end{listing}

In this implementation, an event listener is a boxed function consuming \mintinline{rust}|&dyn Any|, the Rust trait for dynamic typing. An event registry is a hashmap from strings to vectors of listeners. However, this implementation shifts significant complexity and risk onto the API client. \Cref{lst:event-string-ex} shows an example client usage. Observe that the event listener must assert that the event payload is typed as \mintinline{rust}|OnClick| via the \mintinline{rust}|downcast_ref| method. The \mintinline{rust}|register| and \mintinline{rust}|trigger| calls must both use a matching \mintinline{rust}|"click"| string. The \mintinline{rust}|trigger| call must also provide a payload of the appropriate type. If the API client messes up any of these conditions, a runtime error will occur.

To develop a safer alternative, the key idea is to unify the representation of event-identity in the API. Rather than using strings to talk about types, we instead use the type directly, facilitated with the \mintinline{rust}|TypeId| API that provides a unique, hashable identifier for any type. \Cref{lst:event-type} shows an implementation of \mintinline{rust}|Events| that maps from type IDs to type-erased vectors of listeners. The \mintinline{rust}|register| and \mintinline{rust}|trigger| methods no longer take a string event name as input, but instead use the generic type \mintinline{rust}|E| to refer to events.

\begin{listing}[h!]
\begin{minted}{rust}
trait Listener<E> = Fn(&E) -> () + 'static;
type ListenerVec<E> = Vec<Box<dyn Listener<E>>>;

#[derive(Default)]
struct Events {
  listeners: HashMap<TypeId, Box<dyn Any>>
}

impl Events {
  fn register<E: 'static, F: Listener<E>>(
    &mut self, f: F
  ) {
    let fs = self.listeners
      .entry(TypeId::of::<ListenerVec<E>>())
      .or_insert_with(|| 
        Box::new(Vec::<ListenerVec<E>>::new()));    
    fs.downcast_mut::<ListenerVec<E>>().unwrap()
      .push(Box::new(f));
  }

  fn trigger<E: 'static>(&self, ev: &E) {
    let fs_opt = self.listeners
      .get(&TypeId::of::<ListenerVec<E>>());
    if let Some(fs) = fs_opt {
      let fs = fs.downcast_ref::<ListenerVec<E>>()
        .unwrap();
      for f in fs { f(ev); }
    }
  }
}    
\end{minted}
\vspace{-1em}
\caption{Event registry with type-key encoding.}
\label{lst:event-type}
\end{listing}

\Cref{lst:event-type-ex} shows an example usage of this alternative API. Observe that all the previous failure points are gone. A callback's payload must be consistent with the associated event. A triggered event's payload must also be consistent with the associated event. Any typos are caught by the compiler.

\begin{listing}[h!]
\begin{minted}{rust}
struct OnClick { mouse_x: usize, mouse_y: usize }
let mut events = Events::default();

events.register(|ev: &OnClick| {
  assert_eq!(ev.mouse_x, 1);
});
let ev = OnClick { mouse_x: 1, mouse_y: 3 };
events.trigger(&ev);

// Can't associate this with OnClick
events.register(|ev: &OnKeyPress| { /* ... */ });

// Only triggers OnKeyPress callbacks
events.trigger(&OnKeyPress { /* ... */ });

// Typos caught by the compiler
events.trigger(&OnClack { /* ... */ });
\end{minted}
\vspace{-1em}
\caption{Example usage of the type-key event encoding.}
\label{lst:event-type-ex}
\end{listing}

\end{patpart}

\begin{patpart}{Commentary}
The string-key event system is an instance of the broader problem of ``stringly-typed'' systems. For example, most developers learn early on that a fixed domain of objects is better represented by an enumerated type rather than a string. But that lesson can be difficult to extrapolate into open-world domains like an event system. Typed design patterns like the Registry can highlight the mechanisms needed in such a case, like the \mintinline{rust}|TypeId| API.

In the Rust ecosystem, typed registries can be found in many crates. The Matrix client for Rust\,\citegithub{https://github.com/matrix-org/matrix-rust-sdk} uses a typed registry for event handling. The Bevy game engine\,\citeurl{https://bevyengine.org/} uses a typed registry for implementing a typed entity-component-system architecture. The Shaku library\,\citegithub{https://github.com/AzureMarker/shaku} provides a generic dependency injection framework using typed registries. An implementation of the Liquid template language in Rust\,\citegithub{https://github.com/cobalt-org/liquid-rust} uses typed registries to hold external plugins. The core data structure that maps type IDs to boxed values can be found neatly encapsulated in the AnyMap library\,\citegithub{https://github.com/chris-morgan/anymap}.
\end{patpart}

\section{Related Work}
\label{sec:rw}

Having now seen four patterns in full detail, we should step back and consider how this style of presentation relates to prior work on teaching functional system design. Several books articulate design patterns for individual functional languages: \textit{Scala Design Patterns}\,\cite{hunt2013scala}, \textit{Functional Programming Patterns in Scala and Clojure}\,\cite{bevilacqua2013scala}, \textit{Scala Functional Programming Patterns}\,\cite{khot2015scala}, \textit{Haskell Design Patterns}\,\cite{lemmer2015haskell},  \textit{Scala Design Patterns}\,\cite{nikolov2016scala}, and \textit{Functional Design and Architecture}\,\cite{granin2020haskell}.

These books do not articulate design patterns in the same sense I use within this paper. For example, much of the content of these books is (a) reimplementing GoF patterns like visitors, (b) explaining standard library APIs like monoids and functors, or (c) explaining basic functional programming concepts like currying. That content is assuredly valuable to its readers, but I do not envision that the $N^{\text{th}}$ tutorial on monads should be considered a functional design pattern.

My philosophy is more ideologically aligned with books that articulate a modeling-oriented design practice. For example, \textit{How to Design Programs}\,\cite{felleisen2018design}, \textit{Domain Modeling Made Functional}\,\cite{wlaschin2017ddd}, and \textit{Type-Driven Development with Idris}\,\cite{brady2017idris} all fall under this umbrella. The difference between these books and the present work is, to an extent, stylistic. These books present a monolithic narrative that is intended to be consumed linearly from the first to the last page. But this style is less well-suited to demand-driven learning where a developer may want to simply find the best patterns for the problem at hand.

\section{Discussion}

One goal of this paper is to advance my particular vision for functional design patterns. But another is to prompt a discussion within the relevant academic and industrial circles about the key question: what kinds of knowledge, if any, ought to be articulated as typed functional design patterns? Or put another way, what do developers need to know that isn't currently documented about how to design effective systems in functional languages?

To that end, we should briefly consider the negative space of issues unaddressed by this paper.  For instance, this paper has focused primarily on design patterns that catch incorrect programs at compile-time. But performance is also an important aspect of system design. Within the Rust community, there are excellent resources for how to profile and optimize small pieces of code like \textit{The Rust Performance Book}\,\cite{rust-perf}. However, there are comparatively few resources for how to architect a system to trade-off correctness, performance, maintainability, compile-times, etc. at larger scales. As just one example of a performance pattern, many Rust libraries use interning strategies (with a range of implementations) to reduce the memory footprint of commonly-allocated objects. Such performance patterns deserve to be documented in their own right.

Regardless, I believe it is high time to revisit design patterns now that the functional era is upon us. I hope to be a part of writing \textit{Functional Design Patterns}, and I invite you to join me in making a Gang of more than just One.

\bibliographystyle{ACM-Reference-Format}

\end{document}